\documentstyle[preprint,aps]{revtex}
\begin{document}

\draft

\title{Magnetic coupling in 
mesoscopic metal/ferromagnet layered systems}

\author{B.Spivak}

\address{Physics Department, University of Washington, Seattle, WA 98195, 
USA and Max-Plank-Institut fur Festkorperforschung Hochfeld-Magnetolabor,
Grenoble, France}

\author{A.Zyuzin}

\address{A.F.Ioffe Physical- Technical Institute, 194021, St.Petersburg, 
Russia}

\maketitle

\begin{abstract}
We consider a mesoscopic mechanism of the exchange interaction
 in a system of alternating ferromagnetic/nonmagnetic metallic layers.  
In the case of small mesoscopic samples the sign and the amplitude of the 
the exchange interaction energy turn out to be random sample-specific 
quantities. 
They can be changed by applying an external magnetic field, by attaching 
to the system superconducting electrodes with different phases of 
the superconducting order parameter and by changing the chemical 
potential of electrons in the metal with the help of a gate. In the case 
of square or cubic geometries of the nonmagnetic layer at low temperature 
the variance of the exchange energy turns out to be sample size 
independent.  
 \end{abstract}


\newpage

The investigation of oscillations of the exchange energy
 as a function of the nonmagnetic layers' thickness and the "giant 
magnetoresistance" in the system of
 alternating ferromagnetic/nonmagnetic metallic layers has both 
scientific and technological interest $^{[1-10]}$. 

The origin of the exchange interaction between the ferromagnetic (F) 
layers is the Ruderman- Kittel type interaction: The interaction of
itinerant nonmagnetic metal electrons with localized ''f'' or ''d'' 
electrons in 
the ferromagnets induces a spin polarization in the nonmagnetic 
metallic (N)
layers. This magnetization, in turn, creates the effective interaction
 between two localized spins in different ferromagnetic layers with the 
energy 
\begin{equation}
E(\vec{R}_{1}, \vec{R}_{2})=I_{ij}(\vec{R}_{1}, 
\vec{R}_{2})S^{1}_{i}S^{2}_{j}.
\end{equation}  
Here $S^{1,2}_i$ are components of localized spins in the F 
layers and $\vec{R}_{1,2}$ are their coordinates.
We employ the simplest model where conduction "s"-electrons 
interact with localized "f" or "d" electrons in the F layers via a 
contact interaction with energy 
\begin{equation}
A\sum_{k}\delta(\vec{r}_{k}-\vec{R})\vec{s}_{k}\vec{S}.
\end{equation}
Here $\vec{r}_{k}$ and $\vec{s}_{k}$ are 
coordinates and spins of conduction electrons in the metal which are labeled by the index $k$ and $A$ is the 
interaction constant.
In the case where two magnetic ions are embedded in a 3-d nonmagnetic
 metal Eqs.1,2  lead to a well known form of the exchange energy $^{[11]}$
\begin{equation}
I_{ij}(\vec{R}_{1}, \vec{R}_{2})= 
I_{0}\frac{\cos(2p_{F}|\vec{R}_{1}-\vec{R}_{2}|)}{(p_{F}|\vec{R}_{1}-
\vec{R}_{2}|)^3}.
\end{equation}
Here $I_{0}=\frac{9\pi}{64}\frac{(An)^2}{E_{F}}$ is the
interaction energy of 
adjacent localized spins in ferromagnets, which is of order of 
ferromagnet's critical temperature; $E_{F}$, $p_{F}$ are the Fermi energy 
and the Fermi momentum, respectively, and $n=\frac{p_{F}^{3}}{3\pi^2}$ is 
the concentration of electrons. Following to 
$^{[6,7,8,9]}$ we will use the approximation where the total exchange 
interaction 
energy $\bar{E}$ between the magnetic moments in the F layers is a 
sum of $E(\vec{R}_{1}, \vec{R}_{2})$  over coordinates $\vec{R}_{1}$, 
$\vec{R}_{2}$ of the localized spins in the 
ferromagnetic layers.
\begin{equation}
\bar{E}=\sum_{\vec{R}_{1},\vec{R}_{2}}I_{ij}(\vec{R}_{1}, 
\vec{R}_{2})S_{i}^{(1)}S_{j}^{(2)}=\bar{I}_{ij}n_{1i}n_{2j} 
\end{equation}
Here $n_{1i}$ and $n_{2j}$ are components of unit vectors $\vec{n}_{1}$, $\vec{n}_{2}$ parallel to 
magnetizations in the first and the second layers respectively. We assume 
that 
the exchange energy inside ferromagnets is large enough and fluctuations
of spin magnetization direction along the F layers can be neglected. As a 
result, in the case of 
clean N layers ($L \ll l$), the value of $\bar{I}_{ij}$ and the relative 
orientation of magnetizations of different F layers are oscillating 
functions of $L$. $^{[4-9]}$ Here $L$ and $l$ are the thickness
 and the elastic electron mean free path the in N layer, respectively.
The same effects takes place in the system of alternating 
ferromagnetic/degenerated semiconductor layers. 

Until now, however, both experimental and theoretical studies of
this phenomenon have been restricted to the consideration of the 
infinite dimensions of both F and N layers and clean N layers.
In this article we discuss the opposite case of small sample sizes 
and disordered N layers where the mesoscopic
effects determine both the exchange interaction between the ferromagnetic 
layers and the conductance of the system.
Let us consider, for example, a system of two F layers of sizes $L_{1}, 
L_{2}, L_{3}$ divided by 
a disordered N layer with $L>l$ (See Fig.1).
 In this case the amplitude 
of the Ruderman-Kittel
exchange interaction between a couple of spins located inside different 
F-layers $\langle I_{ij}(\vec{R}_{1}, \vec{R}_{2})\rangle \sim 
\exp(-\frac{|\vec{R}_{1}-\vec{R}_{2}|}{l})\sim \exp(-\frac{L}{l})$, 
averaged over realizations of a random potential (and consequently the total 
average exchange energy $\langle\bar{E}\rangle$ between F layers) is
exponentially small and can be neglected. $^{[12]}$ Here brackets $\langle \rangle$ 
correspond to averaging over realizations of a random scattering 
potential. On the other hand, 
it is well known $^{[13-16]}$ that the exponential decay of the average 
$\langle I_{ij}(\vec{R}_{1}, \vec{R}_{2})\rangle$ is the
consequence of randomization of the sign of $I_{ij}(\vec{R}_{1}, 
\vec{R}_{2})$ and that the typical amplitude
of the interaction $\sqrt{\langle (I_{ij}(\vec{R}_{1}, \vec{R}_{2}))^2\rangle}\sim 
|\vec{R}_{1}-\vec{R}_{2}|^{-d}$ 
decreases with distance in the same way as in the pure 
case. Here $d$ is the dimensionality of space. 
Therefore, in the case $L \gg l$ the amplitude of the Ruderman-Kittel 
interaction between the ferromagnetic layers must be controlled by the 
mesoscopic effects. 

In this article we discuss the typical magnitude of the "mesoscopic part" 
of the exchange energy between the ferromagnetic layers. We show that it is 
determined by long range correlations between $I_{ij}(\vec{R}_{1}, 
\vec{R}_{2})$ and $I_{k,l}(\vec{R}_{3}, \vec{R}_{4})$ which survive even 
on distances $|\vec{R}_{1}-\vec{R}_{3}|, |\vec{R}_{2}-\vec{R}_{4}|\gg l$.
As a result, for example, in the case of square or cubic geometries of the 
N layer and at low temperatures the variance of the exchange energy turns 
out to be sample size independent.
 
Let us start with the case where the length $L_{2}$ of the F-layers is 
relatively short and one can neglect the fluctuations of the orientation 
of magnetizations along the F layers. To find the
relative angle $\theta(\widehat{\vec{n}_{1},\vec{n}_{2}})$ between 
magnetization's angles in different layers one has to
calculate the sign and the amplitude of the quantity $\bar{I}_{ij}$.
In the case where metallic region is disordered, $\bar{I}_{ij}$ is a random 
sample-specific quantity, which can be characterized by its average and
 moments. One can calculate the correlation function 
$\langle \delta\bar{I}_{ij}\delta\bar{I}_{kl}\rangle$ with the help of the standard 
diagram
technique $^{[17]}$ for averaging over realizations of random potential. 
 Here
$\delta \bar{I}_{ij}=\bar{I}_{ij}-\langle \bar{I}_{ij}\rangle$.
The diagrams which contribute to $\langle \delta\bar{I}_{ij}\delta\bar{I}_{kl}\rangle$ to 
lowest order in parameter $\frac{\hbar}{p_{F}l}\ll 1$
are shown in Fig.2,  where solid lines correspond to electron Green 
functions in Matsubara representation, dashed lines correspond to the 
scattering on the random potential
and vertexes correspond to the contact magnetic interaction Eq.2.
The blocks of the diagrams shown in Fig.2c  correspond to the so called 
Cooperons and Diffusons $\hat{P}^{c,d}_{\omega}(\vec r, \vec r')$, which 
are tensors as functions of electron spin indexes $^{[17]}$.
The diagrams shown in Fig.2a have been considered in $^{[13-16]}$. They give 
main contribution to the correlation function
$\langle\delta I_{ij}(\vec{R}_{1},\vec{R}_{2})\delta I_{kl}(\vec{R}_{1}, 
\vec{R}_{2})\rangle\sim |\vec{R}_{1}-\vec{R}_{2}|^{-d}$. These diagrams 
correspond, however, to the correlation
function $\langle\delta I_{ij}(\vec{R}_{1}, \vec{R}_{2})\delta I_{kl}(\vec{R}_{3}, 
\vec{R}_{4})\rangle$ which 
 decays
exponentially when $|\vec{R}_{1}-\vec{R}_{3}|, |\vec{R}_{2}-\vec{R}_{4}|\gg l$.
As a result, for example, in the case $L\sim L_{2}\gg L_{3}\sim L_{1}$ the 
contribution to 
$\langle\delta\bar{I}_{ij}\delta\bar{I}_{kl}\rangle$ from these diagrams 
is of the order of 
\begin{equation}
I^{2}_{0}(n p_{F}^{-3})^{4} (\frac{L_{1}}{L})^{2}.
\end{equation}
We assume the density of localized spins in the ferromagnets in of order 
of $n$.

 The diagrams Fig.2b  give much smaller contribution to 
$<\delta I_{ij}(\vec{R}_{1}, \vec{R}_{2}))\delta I_{kl}(\vec{R}_{1}, 
\vec{R}_{2})>$, 
but they describe long range correlations $\langle\delta I_{ij}(\vec{R}_{1}, 
\vec{R}_{2})\delta I_{ij}(\vec{R}_{3}, \vec{R}_{4})\rangle\sim R^{-4(d-1)}$ 
when 
$|\vec{R}_{1}-\vec{R}_{3}|\sim |\vec{R}_{2}-\vec{R}_{4}|=R\gg l$.
As a result, it is these diagrams that determine the 
of the correlation function of the interlayer exchange energy 
$\langle\bar{I}_{ij}\bar{I}_{kl}\rangle$ at $L\gg l$.
 The qualitative explanation of the origin of the correlation is as follows. The mesoscopic
fluctuations of the exchange energy $\delta I_{ij}(\vec{R}_{1}, 
\vec{R}_{2})$ result from
 the interference of random probability amplitudes of
diffusion paths between the points $\vec{R}_{1}$ and $\vec{R}_{2}$. Among 
these paths there are some which visit points $\vec{R}_{3}$ and 
$\vec{R}_{4}$ (An example is the line "a" in Fig.1a).  This leads to the 
mentioned above correlation between
$I_{ij}(\vec{R}_{1}, \vec{R}_{2})$ and $I_{kl}(\vec{R}_{3}, \vec{R}_{4})$.
As a result, we have: 
\begin{eqnarray}
\langle\delta\bar{I}_{ij}\delta\bar{I}_{kl}\rangle 
=\frac{2}{\pi}I^{2}_{0}E_{F}^{2} T\sum_{m}\omega\int 
d\vec{R}_{1} d\vec{R}_{2} d\vec{R}_{3}d\vec{R}_{4}  
\nonumber \\
(\hat{\sigma}_{i}\hat{P}^{c}_{\omega}(\vec{R}_{1}, 
\vec{R}_{2})\hat{\sigma}_{k}\hat{P}^{c}_{\omega}(\vec{R}_{2}, \vec{R}_{3}) 
\hat{\sigma}_{j}\hat{P}^{c}_{\omega}(\vec{R}_{3}, 
\vec{R}_{4})\hat{\sigma}_{l}\hat{P}^{c}_{\omega}(\vec{R}_{4}, 
\vec{R}_{1})+ \nonumber \\
 \hat{\sigma}_{i}\hat{P}^{d}_{\omega}(\vec{R}_{1}, 
\vec{R}_{2})\hat{\sigma}_{j}\hat{P}^{d}_{\omega}(\vec{R}_{2}, 
\vec{R}_{3})\hat{\sigma}_{k}\hat{P}^{d}_{\omega}(\vec{R}_{3}, 
\vec{R}_{4})\hat{\sigma}_{l}\hat{P}^{d}_{\omega}(\vec{R}_{4}, \vec{R}_{1})). 
\end{eqnarray}
Here $\omega=\pi (2m+1)T$ is the Matsubara frequency, $m$ is an integer 
number $T$ is the temperature and $\hat{\sigma}_{i}$ are spin operators.
Integrations over $\vec{R}_{1}$, $\vec{R}_{3}$ and $\vec{R}_{2}$, 
$\vec{R}_{4}$
in Eq.6 is performed over volumes of the first and the second ferromagnetic 
layers respectively. 
Results of calculation of Eq.6 depend on the ratio between the lengths 
$L$, $L_{2}$, $L_{T}=\sqrt{\frac{D}{T}}$,   
$L_{so}=\sqrt{D\tau_{so}}$ and on boundary conditions for 
Cooperons and Diffusons. Here $L_{so}$, $\tau_{so}$ are the spin-orbit 
relaxation length and time, respectively, and $D$ is the electron diffusion 
coefficient in the N layer. In the case of 
the "open" geometry of  the N layer shown in Fig.1a and $L_{T}, 
L_{so}\gg L>l$; $L, L_{2}\gg L_{1}, L_{3}$, we have

\begin{equation}
<\delta\bar{I}_{ij}\delta\bar{I}_{kl}> 
=\frac{5\cdot 
2^{\frac{7}{2}}\zeta(\frac{5}{2})}{3^{2}\pi^{\frac{9}{2}}}X\frac{I_{0}^{2}}{(p_{F}l)^2}(p_{F}L_{1})^{4}\delta_{ij}\delta_{kl} \end{equation}
Here $X$ is a factor, which is of the order of unity when $L\sim 
L_{2}\leq L_T$ and $\zeta(x)$ is the Zeta function. In different limiting 
cases we have:

\begin{equation}
X=\left \{ \begin{array}{cc}
(\frac{L_2}{L})^{4}
  &
\mbox{if $L_{T}> L_{2}> L$} \\
\frac{L_{2}L_{T}^{3}}{L^4} & \mbox{if $L_{2}>L_{T}>L$}
\end{array}
\right.
\end{equation}

In the case $L>L_{T}$ the expression for $X$ acquires an additional 
exponentially small factor $\exp(-\frac{L}{L_T})$.
In the case $L_{so}>L$ the minimum of the exchange energy corresponds to 
parallel or antiparallel orientation of the layer's magnetizations 
($\theta$ equals zero or $\pi$).
 In the 
opposite limit $L_{so}\ll L$ we get the same formula as Eq.8  but without the 
factor $\delta_{ij}\delta_{kl}$.
This means that the exchange interaction between the F layers is of the 
Dzialoshinski-Moria type and a minimum of the exchange energy corresponds 
to a sample specific angle $\theta(\widehat{\vec{n}_{1},\vec{n}_{2}})$ 
distributed randomly in the interval $(0,\pi)$.
It is interesting that in the case $L\sim L_{2}<L_T$
Eq.7 turns out to be independent of $L$. Let us note that the diagrams 
shown in Fig.2a  give $\sqrt{\langle(\delta \bar{I}_{ij})^{2}\rangle}\sim 
L^{-1}$.

While deriving the results presented above we neglected the sensitivity of
the boundary
conditions for Cooperons and Diffusons to the change of magnetization
directions in F-layers. In the case of the open sample geometry Fig.1a
this is correct, provided $\frac{Ap^{3}_{F}L_{1}}{v_{F}}\ll 1$.  To get an
estimate for $\langle\delta\bar{I}_{ij}\delta\bar{I}_{kl}\rangle$ in the
opposite limit one has to substitute the factor $A(L_{1}p_{F})$ in Eq.8 for
$E_{F}$. For example, in the case $L_{T}>L\sim L_{2}>L_{so}$ we have
\begin{equation}
\langle\delta\bar{I}_{ij}\delta\bar{I}_{kl}\rangle\sim
E^{2}_{F}(p_{F}l)^{-2}\sim \frac{\hbar}{\tau}.
 \end{equation}
Here $\tau$ is the elastic mean free path in
the metal.
We would like to stress that the origin of both Eq.7 and Eq.9 is the 
exstance of the long range correlation of signs if the quantities 
$I_{ij}(\vec{R}_{1},\vec{R}_{2})$ and $I_{kl}(\vec{R}_{3},\vec{R}_{4})$ 
on distances much larger than $l$.

In the case when the length of the ferromagnetic layers $L_2$ is long enough
one should take into account the random fluctuations of the direction of
magnetization along the layers. A solution of this problem is beyond the
scope of this paper.

As it is usual in the physics of mesoscopic metals $^{[19,20]}$, the 
external 
magnetic field changes the electron interference pattern and, 
thereby,
$\theta(\widehat{\vec{n}_{1}\vec{n}_{2}})$ turns out to be a random 
sample-specific oscillating function of the magnetic field 
$H$. In the case of the open geometry of the N layer 
shown in Fig.1a for $L_{2}\sim L$ the characteristic period of these 
fluctuations 
is $^{[19,20]}$ $\delta H_{1}\sim 
\frac{\Phi_0}{L^2}$. Here $\Phi_0$ is the flux quantum. 
There is also another characteristic magnetic field 
in the system which corresponds to the interaction energy between the 
external magnetic 
field and magnetic moments of the F layers which is of order of the exchange 
energy between the layers: $\delta H_{2}\sim 
\frac{\bar{I}_{ij}}{\mu p^{3}_{F}L_{1}L_{2}L_{3}}\sim 
\frac{\hbar}{\tau}\frac{1}{\mu p^{3}_{F}LL_{1}L_{3}}$. Here $\mu$ is 
the Bohr magneton.  
 At large enough $L$, $\delta H_2>>\delta H_1$ and
therefore the exchange 
energy $\bar{I}_{ij}$ will be a random sample-specific oscillating 
function of the magnetic field in
the situation where the interaction energy between the magnetic field and the
ferromagnetic moment is still negligible.
As a result, the relative orientation of magnetizations of the
F-layers, characterized 
by $\theta(\widehat{\vec{n}_{1},\vec{n}_{2}})$, will 
be a random function of the magnetic field. In the opposite limit, when 
$\delta H_{2}<\delta H_{1}$, $\theta(\widehat{\vec{n}_{1},\vec{n}_{2}})$ 
monotonically decreases with the magnetic field.
Even in this 
case it is possible to see the random oscillations of 
$\theta(\widehat{\vec{n}_{1},\vec{n}_{2}})$ near the magnetic field 
corresponding to spin-flop transitions.

Another way to change the relative orientations of the F-layers 
is demonstrated in Fig.1b. Namely, 
$\theta(\widehat{\vec{n}_{1},\vec{n}_{2}})$ is a random sample specific
function of the order parameter phase difference $(\chi_{1}-\chi_{2})$
in superconductors $S_1$ and $S_2$ in Fig.1b. The reason for this is
that some diffusive paths connecting points 1 and 2 in Fig.1b can visit
superconductors (line "b" in Fig.1b) and the corresponding amplitude of
probability to travel along these paths acquire the additional phase 
$(\chi_{1}-\chi_{2})$ $^{[20]}$. Another consequency of the phase 
dependence of the exchange energy is that the critical Josephson current 
of the device shown in Fig.1b depends on the angle $\theta$ between 
magnetizations of F layers. At last, a change of electrical or 
chemical potentials in the metal (or in the degenerate semiconductor)
 also lead to the oscillations of $\bar{I}_{ij}$.
In the case $L\sim L_2$ characteristic period of the oscillations as a 
function of 
the chemical potential $\mu$ is of the order of $\delta\mu\sim 
E_{c}=\frac{D}{L^{2}}$. $^{[19]}$
In the case when the degenerate semiconductor is a part of a MOSFET one 
can change $\mu$ by changing the voltage on the gate.

The resistance of the system considered above is a
random oscillating function of the external magnetic field. There are two 
mechanisms which
cause these oscillations. 1.The usual mechanism for mesoscopic metallic
samples: $^{[19,20]}$ amplitudes of probability to travel along 
diffusion
paths  acquire additional random phases, proportional to the magnetic field.
The characteristic period of the oscillations is of the order of $\delta
H_1$. 2.The magnetic field induced change of 
$\theta(\widehat{\vec{n}_{1},\vec{n}_{2}})$ which corresponds to a change of 
the boundary conditions
for electrons in nonmagnetic metal.As a result the rotation of the 
ferromagnetic layer's magnetizations leads to mesoscopic fluctuations of 
the resistance of the sample. The characteristic 
period of the oscillations correspond to $(\theta(H)-\theta(0))\sim 1$. 

 The calculations presented above do not take into account 
the dipole interaction
between magnetic moments in different F-layers. This interaction decays 
with the distance between the layers $L$ and can be much less than the 
considered above mesoscopic part of the exchange energy, which is 
independent of $L$ if $L\ll L_{T}$.

In conclusion, we would like to mention that the considered above mechanism 
can determine random anisotropy of the exchange interaction 
in bulk disordered ferromagnets. Let us introduce an exchange field 

\begin{equation}
\hat{\vec{H}}_{ex}=\frac{1}{v}\int_{v}\vec{H}_{ex} d\vec{r}.
\end{equation} 
 averaged over a volume $v=L_{0}^{3}$, $l<L_{so}<L_{0}<L_T$. Here 
$\vec{H}_{ex}$ is the local 
exchange field acting on localized spins in ferromagnet. The correlation 
function of this field  
can be calculated with the help of the diagram Fig.2b.

\begin{equation}
\langle\hat{H}_{exi}(\vec{r})\hat{H}_{exj}(\vec{r'})\rangle=
\frac{I^{2}_{0}}{\mu^{2}(p_{F}l)^2}\frac{1}{(p_{F}L_0)^2}f(|\vec{r}-\vec{r'}|)
\end{equation}
 Here $f(r)$ is the function which is equal to unity at $r<<L_{0}$, 
decays as 
$f(r)\sim (\frac{L_T}{r})^6$ at $L_{0}<r<L_T$ and exponentially decays when 
$r>L_T$. The above-mentioned long 
range correlation between $\delta I_{ij}(\vec{R}_{1},\vec{R}_{2})$ and 
$\delta I_{kl}(\vec{R}_{3},\vec{R}_{4})$ manifests itself in the factor 
$(p_{F}L_{0})^{-2}$ in Eq.11.
An assumption about short range correlations would lead to an expression 
which is proportional to $(p_{F}L_{0})^{-3}$.

The authors would like to acknowledge useful discussions with 
J.Bass, Y.Shender and P.Levy.
This work was partially supported by Division of Material Sciences, 
U.S.National Science Foundation under Contract No. DMR-9205144.

\figure{Fig.1. Schematic pictures of the ferromagnet-nonmagnet layered 
systems. 
\figure{Fig.2 Diagrams for the correlation function 
$\langle\delta\bar{I}_{ij}\delta\bar{I}_{kl}\rangle$. 
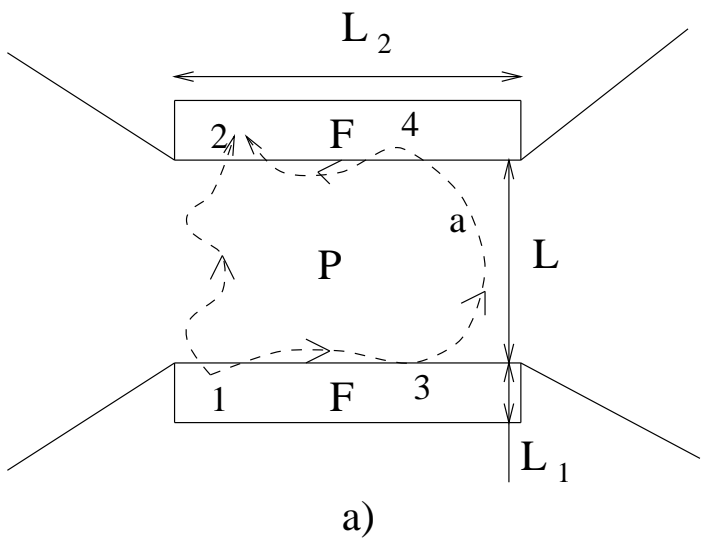
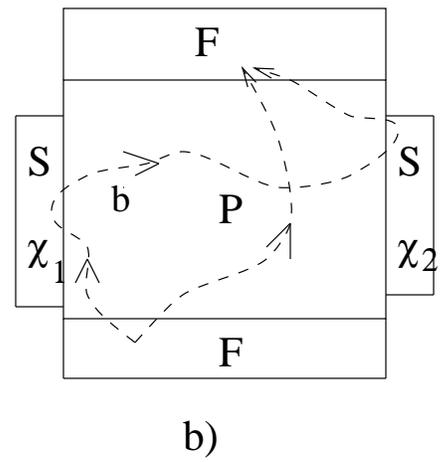
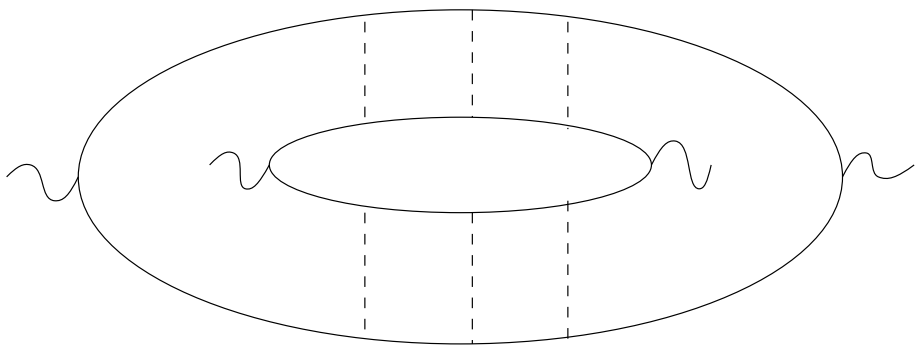
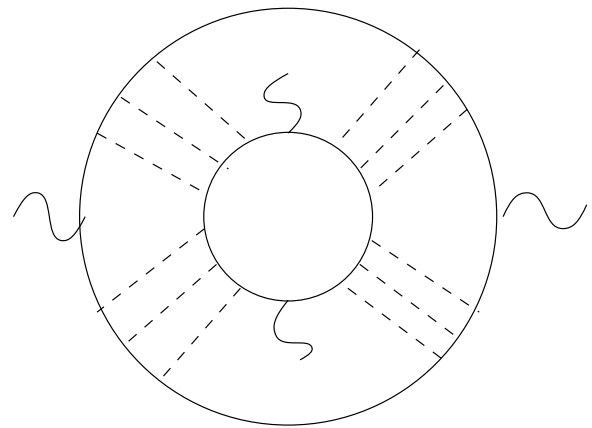
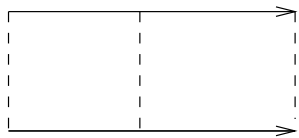
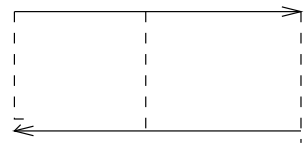
\end{document}